# Controlling the Infrared Dielectric Function through Atomic-Scale Heterostructures

Daniel C. Ratchford[1*], Christopher J. Winta[2*], Ioannis Chatzakis[3], Chase T. Ellis[1], Nikolai C. Passler[2], Jonathan Winterstein[1], Pratibha Dev[4], Ilya Razdolski[2], Joseph G. Tischler[1], Igor Vurgaftman[1], Michael B. Katz[5], Neeraj Nepal[1], Matthew T. Hardy[1], Jordan A. Hachtel[6], Juan Carlos Idrobo[6], Thomas L. Reinecke[1], Alexander J. Giles[1], D. Scott Katzer[1], Nabil D. Bassim[1,7], Rhonda M. Stroud[1], Martin Wolf[2], Alexander Paarmann[2, §], and Joshua D. Caldwell[1,8,§§]

[1] U.S. Naval Research Laboratory, Washington, D.C. 20375, United States

[2] Physikalische Chemie, Fritz-Haber-Institut der MPG, Faradayweg 4-6, 14195 Berlin, Germany

[3] ASEE postdoctoral associate, U.S. Naval Research Laboratory, Washington, D.C. 20375, United States

[4] Department of Physics and Astronomy, Howard University, Washington, DC 20059, United States

[5] NRC postdoctoral associate, U.S. Naval Research Laboratory, Washington, D.C. 20375, United States

[6] Center for Nanophase Materials Science, Oak Ridge National Laboratory, Oak Ridge, Tennessee 37831, USA

[7] Department of Materials Science and Engineering, JHE 357, McMaster University, Hamilton, Ontario, Canada

[8] Department of Mechanical Engineering, Vanderbilt University, 2400 Highland Ave, Nashville, TN 37212, United States

§ Correspondence should be addressed to josh.caldwell@vanderbilt.edu or alexander.paarmann@fhi-berlin.mpg.de

* Contributed equally to this work.

**Abstract**

Surface phonon polaritons (SPhPs) – the surface-bound electromagnetic modes of a polar material resulting from the coupling of light with optic phonons – offer immense technological opportunities for nanophotonics in the infrared (IR) spectral region. Here, we present a novel approach to overcome the major limitation of SPhPs, namely the narrow, material-specific spectral range where SPhPs can be supported, called the Reststrahlen band. We use an atomic-scale superlattice (SL) of two polar semiconductors, GaN and AlN, to create a hybrid material featuring layer thickness-tunable optic phonon modes. As the IR dielectric function is governed by the optic phonon behavior, such control provides a means to create a new dielectric function distinct from either constituent material and to tune the range over which SPhPs can be supported. This work offers the first glimpse of the guiding principles governing the degree to which the dielectric function can be designed using this approach.

The mid-infrared (MIR) to terahertz (THz) spectral region is of great technological importance to nanophotonics and offers numerous potential applications, including super-resolution imaging,[1-4] enhanced spectroscopy[5] and free-space signaling[6]. For instance, nanophotonic approaches are highly suitable to sense trace levels of chemical species through enhanced vibrational fingerprinting[7] due to molecular vibrational frequencies in this spectral region. One promising avenue for realizing these technologies is to use surface phonon polaritons (SPhPs), which greatly enhance light-matter interactions by confining light to sub-diffractional dimensions at the surface of a polar material[4,8-14]. SPhPs are supported between the transverse (TO) and longitudinal optic (LO) phonon frequencies of polar dielectrics, i.e., in the so-called Reststrahlen band, through the coupling of incident light with the oscillating ionic charges on the polar lattice, resulting in a reflectivity approaching 100%[10]. One significant benefit of SPhPs is the substantially reduced losses they exhibit in comparison to surface plasmon polaritons (SPPs). The fast scattering of electrons in metals and semiconductors leads to these larger optical losses for SPPs, while the much longer scattering time for phonons results in lower losses for SPhPs[15]. Although the intrinsic optical losses associated with SPPs have limited their broad application in many practical devices, SPhPs offer a promising alternative[10,16].

One of the most important technological hurdles for the implementation of SPhPs in nanophotonic and metamaterial technologies is that individual polar materials have a narrow, material-specific spectral range over which SPhPs are supported[17]. Thus, while there are many polar materials that occur in nature featuring Reststrahlen bands that combined cover the entire MIR to THz spectral domain, any given material will only support SPhPs within its own relatively narrow specific band. In many instances, it is necessary to combine other material properties (e.g., an accessible band-gap for active tuning[18] or ferroelectric response) with the polaritonic response at a given frequency. For instance, while AlN offers a Reststrahlen band that overlaps with the 8-12 µm atmospheric window, its ~6 eV band gap implies that free carrier-based tuning methods[18] will be unrealistic for the phonon modes originating in AlN. Therefore, it would be ideal if one could tailor the IR dielectric function of the material, from which the nanophotonic response is derived, while maintaining and combining other functional properties.

**Atomic-scale semiconductor superlattices for tunable infrared photonics**

Several recent experiments have demonstrated the spectral tuning of SPhP resonances by overcoating the material with a functional layer (e.g. phase change material)[19,20], directly modifying the polar semiconductor permittivity through carrier photoinjection[18,21] or by means of gate-bias control in planar heterostructures at the nanoscale[22-24]. However, additional optical losses due to the fast electron scattering rates may be introduced, thereby diminishing the intrinsic advantages of the SPhPs for IR nanophotonics.

Here, we experimentally demonstrate a complementary approach for IR-specific tuning using atomic-scale superlattices (SLs) comprised of commercially established polar semiconductors. When polar materials are combined into SLs with layers a few atoms thick, the SL vibrational modes are modified from that of the bulk phonons of the constituent materials by interfacial chemical bonding, electrostatic effects, and changes to the material lattice constants[25,26]. Therefore, the SL behaves as a new, continuous material with a tunable IR dielectric function that depends on the constituent layer thicknesses. These SL structures are referred to as crystalline hybrids (XH)[17,27] since it is the changes to the chemical structure of the crystal lattice that modify the IR response. We demonstrate the XH approach by using an atomic-scale AlN/GaN SL, which has a highly anisotropic IR response featuring multiple Reststrahlen bands. We show that the phonon modes, which define the upper and lower limits of the Reststrahlen bands, shift by >10 cm$^{-1}$ based on the SL layer thicknesses. This XH approach provides the opportunity to control the optic phonon frequencies, the dispersion of the optical constants, and the bandwidth of the Reststrahlen bands. Therefore, one can create new engineered materials for IR and THz nanophotonics and optoelectronics, incorporating additional functionalities through appropriate material selection, while also still taking advantage of the SPhPs' relatively low losses.

**AlN/GaN superlattice as a Crystalline Hybrid**

For the purposes of demonstrating the XH concept, multiple AlN/GaN SL samples, each with layer thicknesses of ~10 or fewer monolayers, were grown using molecular beam epitaxy (MBE). A representative schematic of the layered sample configuration is provided in Fig. 1a for an AlN/GaN SL grown on a SiC semi-insulating substrate. Wurtzite GaN and AlN are both birefringent with $\omega_{TO,\parallel}$ and $\omega_{LO,\parallel}$ phonon modes that oscillate parallel to the crystal c-axis and $\omega_{TO,\perp}$ and $\omega_{LO,\perp}$ phonon modes that oscillate perpendicular to the c-axis. For GaN (AlN), these phonon frequencies[28,29] are $\omega_{TO,\parallel} = 533\ (614)\ cm^{-1}$, $\omega_{LO,\parallel} = 735\ (893)\ cm^{-1}$, $\omega_{TO,\perp} = 561\ (673)\ cm^{-1}$, and $\omega_{LO,\perp} = 743\ (916)\ cm^{-1}$. In bulk form, AlN and GaN have similar IR responses, with partially overlapping Reststrahlen bands bounded by their respective optic phonon frequencies. This is shown in Fig. 1b, which plots the nominal reflectance of bulk GaN (blue), AlN (orange), and SiC (red).

Two representative structures, referred to as 'A' and 'B', are discussed here. Sample A, shown in the cross-sectional scanning transmission electron microscope (STEM) image in Fig. 1c, consists of 50 alternating layers of AlN and GaN grown on a ~50 nm thick AlN buffer layer. Since this sample was deliberately not rotated during growth, a gradient in the Al- and Ga-flux across the wafer surface resulted in a strong variation in the corresponding layer thicknesses, with values ranging from 2 to 3 nm for the AlN and GaN layers across the wafer. At the location shown in Fig. 1c, each layer is ~ 2 nm thick. Sample B (STEM image in Fig. 1e) contains a SL with 500 atomically thin, alternating layers of AlN and GaN with thicknesses of ~1.2 nm (~4 monolayers) and ~1.4 nm (~5 monolayers), respectively. The layer thicknesses in Sample B are more uniform because the sample was rotated during the growth. The STEM and X-ray diffraction (XRD) measurements imply some chemical intermixing of the layers, and that the SLs are pseudomorphic but with a high degree of chemical segregation (see Supplemental).

The optic phonon modes of SL polar materials[30], including III-nitride materials systems[31-34], have been extensively studied and generally described using both microscopic and macroscopic models[35]. When two materials are combined in a SL, phonon confinement effects can arise, resulting in confined phonon modes that propagate in one material and not the other[25], or new vibrational states can occur that feature phonons propagating in both materials. Interface phonon modes may also be supported, wherein the vibration is localized to the interfaces[35]. For sufficiently thick layers, the resulting confined and interface phonon modes can be accurately described using macroscopic electromagnetic modeling, e.g. the transfer-matrix method, starting from the bulk permittivity of each layer[36]. A particularly simple macroscopic approximation when the SL layers are much thinner than the wavelength of the light is the well-known effective-medium theory for layered structures. However, as the layer thicknesses in the SL are reduced to just a few atomic layers, the SL phonon modes are not well described by conventional effective-medium theory because of the atomic-scale interactions that modify the SL phonon modes, namely, the prominent effects of the interfacial bonds that impose different boundary conditions and the modified lattice constants of the epitaxially grown layers. The distinction between the confined and interface modes is less important in this limit, and we may regard the SL as a new effective XH material. This point is illustrated in Fig. 1d and 1f, where the measured IR reflectance (blue line) is plotted for Samples A and B, respectively, along with the calculated reflectance (orange dotted line) of the SLs using the transfer-matrix method based on the bulk optical constants of AlN and GaN. It is clear that the conventional approach towards predicting the IR reflectance fails, with the quantitative mismatch of peak positions and the emergence of unexpected features in the experimental spectra (especially prominent for the thinner layered Sample B). As a result, the XH opens up additional degrees of freedom in the design of novel SPhP materials that have not been explored until this work.

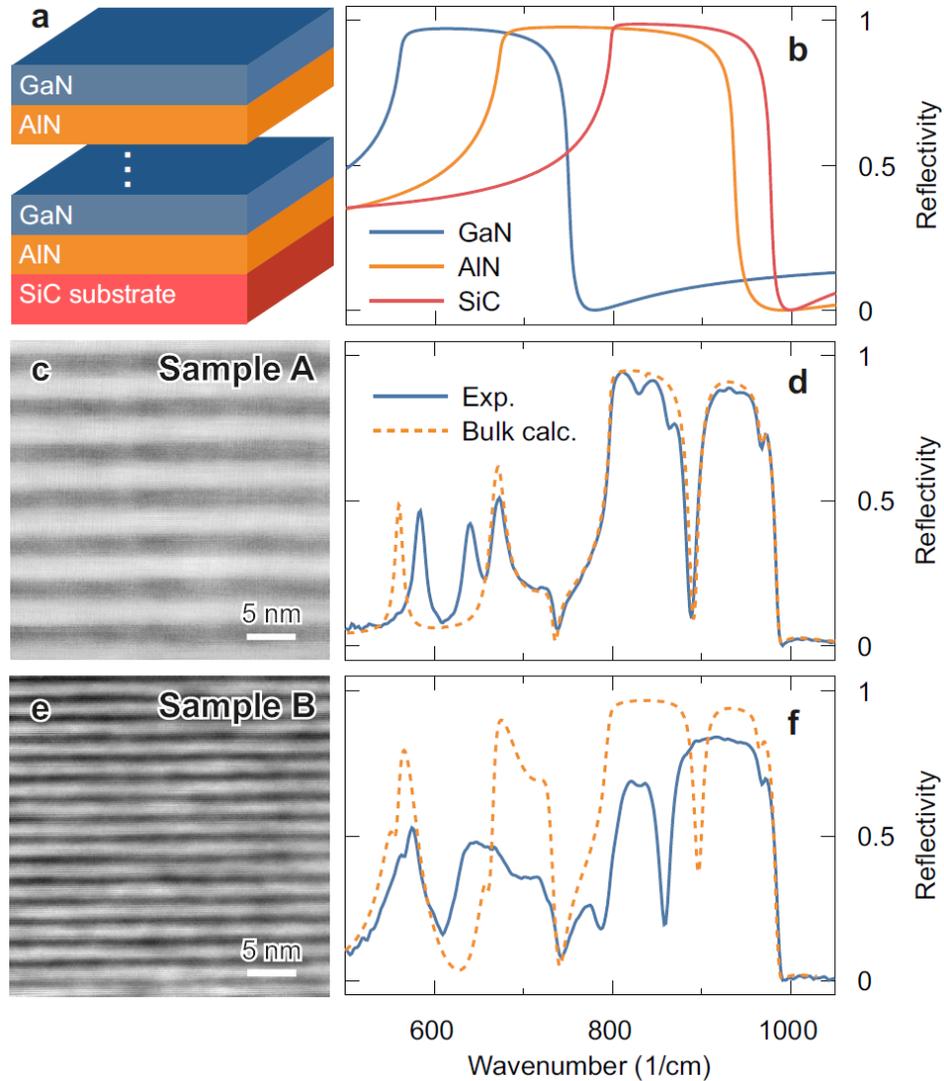

**Figure 1** STEM images and reflectance of AlN/GaN heterostructures and Reststrahlen bands. a) Illustration depicting AlN/GaN heterostructures. b) Calculated reflectance of bulk GaN, AlN, and SiC for normal incidence light in the IR, showing the spectral overlap of the Reststrahlen bands. c) and e) display cross-sectional high-angle annular dark-field STEM images of AlN/GaN heterostructures with AlN (GaN) thicknesses of ~2.2 nm (2.2 nm) for Sample A and ~1.2 nm (1.4 nm) for Sample B, respectively. The AlN layers appear as dark-gray bands and the GaN layers as light-gray bands. d) and f) show the measured reflectance (blue line) at an incidence angle of 65° and the calculation (orange dashed line) based on the bulk properties of the AlN/GaN SL from c) and e), respectively.

**Tunability of the superlattice optic phonons**

We first characterize the optic phonons of the SL, which define the poles and zeros of the dielectric function and the range of the Reststrahlen bands. Quantitative analysis of the phonon modes from the reflection spectra is difficult due to the highly reflective Reststrahlen band of the SiC substrate, along with the high reflectivity of the developing XH Reststrahlen bands between the multiple LO/TO phonon pairs. Thus, in order to directly characterize the SL optic phonons, we employed IR second harmonic generation (SHG),[37-40] collecting spectra from each of the XH structures using an IR free-electron laser[41]. This technique was recently shown to provide clear identification of phonon modes in multimode systems[42] since it allows for the direct measurement of phonon peaks on a nominally flat spectral background.

The IR SHG and reflection spectra collected from three locations on Sample A with varying AlN and GaN layer thicknesses is provided in Fig. 2a and b, respectively. In contrast to the reflection measurements (Fig. 2b), the SHG spectra (Fig. 2a) provides clear, separate peaks that can be analyzed using standard line-shape fitting methods. Within the SHG spectra, each $E_1$(TO) phonon is observed with large contrast[42]. Additionally, $A_1$(LO) phonons of the SL can be closely approximated via the Berreman effect[43] that results in subtle dips in the reflectivity near the spectral position where the real part of the permittivity crosses through zero (so-called epsilon-near-zero condition). This also results in pronounced peaks in the SHG (see Supplemental for details). The spectra in Fig. 2a feature phonon resonances of the SL as well as the SiC substrate and the AlN buffer layer. The latter two are easily identified, as they occur at the bulk TO and LO phonon frequencies and do not vary spectrally across the wafer surface. The SHG peaks originating from the SiC, the AlN buffer, and the SL are highlighted in blue, orange, and green, respectively, in Fig. 2.

From the SHG spectra one can immediately discern that the spectral positions of the SL optic phonons vary strongly across the wafer (see Supplemental for the full dataset and extended discussion). The SL exhibits two $E_1$(TO)-like phonon modes around ~575 cm$^{-1}$ and ~630 cm$^{-1}$, as well as three $A_1$(LO)-like phonons: one at ~735 cm$^{-1}$ and two more at ~840 cm$^{-1}$ and 880 cm$^{-1}$. Surprisingly, we observe significant frequency shifts (> 10 cm$^{-1}$) of these modes even for minute changes in the layer thicknesses, demonstrating the tunability of the IR dielectric function via this XH concept. We expect a significant modification of the lattice constants of the SL, with the in-plane lattice constant being between AlN and GaN and resulting in tensile and compressive strain for the AlN and GaN layers, respectively. In this case, the effective in-plane lattice constant in the structure depends on the thickness ratio between AlN and GaN. This effect, as well as the resulting strain-induced phonon frequency shifts in short-period AlN/GaN SLs have been studied by Paudel and Lambrecht[26]. As we discuss in the Supplemental, the observed phonon frequency shifts are consistent with the modified in-plane lattice constants we measured by X-ray diffraction. Therefore, controlling the AlN/GaN layer thicknesses provides a direct means to tune the phonon frequencies and, thereby, the IR dielectric function. Additionally, we expect phonon confinement and interface phonon formation to become significant at these small layer thicknesses. In our data, the latter is indicated by the emergence of the new $A_1$(LO) modes near ~840 cm$^{-1}$, in excellent agreement with theoretical predictions[26]. Overall, our observations demonstrate the independent tunability of the XH optic phonons through the layer thickness ratio and absolute layer thicknesses of the SL, making use of the modified strain and interface contributions, respectively, to the XH phonon frequencies.

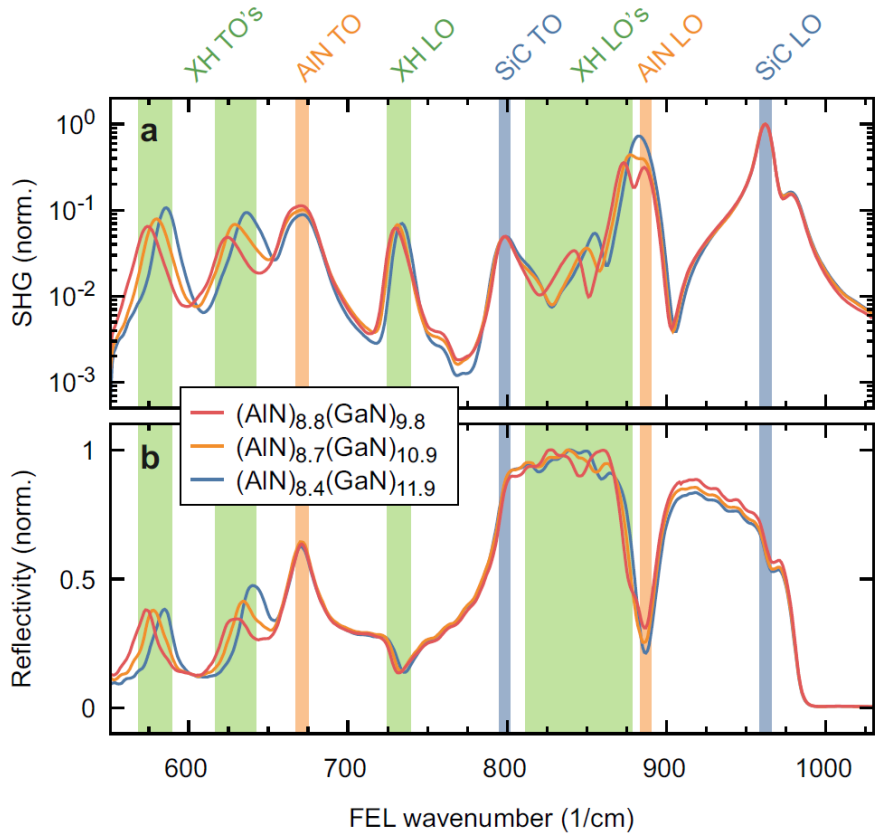

**Figure 2** Optic phonon tunability in the XH structure. (a) Experimental SHG spectra at three exemplary positions on the SL with graded AlN and GaN layer thicknesses (Sample A). Subscripts in the legend denote average number of monolayers. SHG peaks marking the SL's optic phonon modes shift spectrally with varying layer thicknesses, thus demonstrating the tunability of the XH modes (green shade). In contrast, SHG peaks at AlN and SiC bulk phonon frequencies (orange and blue shade, respectively) originating from the buffer layer and the substrate, respectively, do not shift as the SL's layer thicknesses vary. (b) Corresponding reflectivity spectra. The layer thickness-dependent behavior of the XH optic phonons in the linear response is consistent with the features observed in the SHG spectra. Please note the logarithmic scale in (a).

**Dielectric function of the atomic-scale AlN/GaN heterostructure**

We extracted the XH dielectric function of Sample A and Sample B through least-squares fitting of IR ellipsometry data, using a single dielectric function to describe the optical response of each SL sample. The initial fit parameters for the phonon frequencies were taken from the SHG phonon spectra. Here, we focus the discussion on Sample B, which has thinner layers than Sample A and thus maximizes the atomic-scale impact of the XH. Fig. 3a and b shows the measured dielectric function of Sample B. The AlN/GaN XH is strongly birefringent, with very different in-plane and out-of-plane dielectric responses, denoted $\varepsilon_\perp$ (blue curve) and $\varepsilon_\parallel$ (orange curve), respectively. For comparison, Fig. 3c and d shows an effective-medium calculation of the XH dielectric function based on the optical constants for bulk GaN and AlN measured by ellipsometry. From this comparison, it is clear that the effective-medium calculation fails to predict several important features of the IR response because it does not take into account the atomic-scale interactions between the layers.

Perhaps the most direct and powerful means of incorporating the atomistic details (i.e., the effects of the interfacial bonds and modified lattice constants) on the XH phonon modes is through density-functional perturbation theory (DFPT). We employed DFPT to calculate the XH phonon density of states and derive a theoretical dielectric function (Fig. 3e and f). The lattice constants of the SL were assumed to be fully relaxed, and the layer thicknesses were based on Sample B with 5 monolayers of GaN and 4 monolayers of AlN. Unlike the comparison to the effective medium approximation, very good qualitative agreement with experiment is found. The small frequency shifts of the modes between the DFPT and experiment can be attributed to the assumption of fully relaxed lattice constants in each layer as well as the choice of the exchange-correlation functional. In contrast to the effective-medium approach, the DFPT calculation accurately predicts multiple phonon modes at short wavelengths for the out-of-plane permittivity and can provide physical insight into the qualitative vibrational character of the modes.

The measured XH dielectric function exhibits multiple Reststrahlen bands (identified by regions with $Re(\varepsilon) < 0$) as a result of the different phonon modes. The poles in the dielectric function occur at $E_1$(TO)-like and $A_1$(TO)-like phonon frequencies and the zero crossings at $E_1$(LO)-like and $A_1$(LO)-like phonon frequencies for $\varepsilon_\perp$ and $\varepsilon_\parallel$, respectively. The measured $\varepsilon_\perp$ displays two Reststrahlen bands: one narrow region from 572 cm$^{-1}$ to 599 cm$^{-1}$ and another from 629 cm$^{-1}$ to 807 cm$^{-1}$. Therefore, the out-of-plane Reststrahlen bands of the XH cover a broader range than that of GaN (~561-743 cm$^{-1}$) and are on the same scale as AlN (~673-916 cm$^{-1}$). Similar Reststrahlen bands occur in the DFPT dielectric function (Fig. 3e), although they are red shifted by ~15 cm$^{-1}$. Although the strong influence of the atomistic interface conditions in the SL makes unambiguous identification of the modes difficult and less practically useful, the calculated phonon vibrational pattern associated with the 559-583 cm$^{-1}$ Reststrahlen band (Fig. 3g) shows that it most closely resembles a GaN confined mode. In contrast, the phonon associated with the 615-784 cm$^{-1}$ Reststrahlen band (Fig. 3h) appears to be mostly an interface phonon in the AlN layers.

The measured $\varepsilon_\parallel$ is dominated by a broad Reststrahlen band region from about 536 cm$^{-1}$ to 740 cm$^{-1}$. It also exhibits two additional Reststrahlen bands, one from 769 cm$^{-1}$ to 791 cm$^{-1}$ and another from 828 cm$^{-1}$ to 859 cm$^{-1}$, which are present in the DFPT-calculated dielectric function too. As discussed in the Supplemental, examination of the lattice vibrational patterns indicates that these modes involve atomic movement in both the GaN and AlN layers, but in general, these modes are weakly confined to the AlN layers.

The multiple Reststrahlen bands of the XH result in both elliptical and hyperbolic behavior. The former is observed when $\varepsilon_\perp$ and $\varepsilon_\parallel$ have different values, but are both negative in sign, while hyperbolicity results when the permittivities along orthogonal axes are opposite in sign. However, unlike most hyperbolic systems studied to date, the vibrational resonances result in a large range of positive and negative permittivities, both of which vary rapidly with frequency, potentially providing a material system to explore hyperbolic behavior in highly dispersive media.

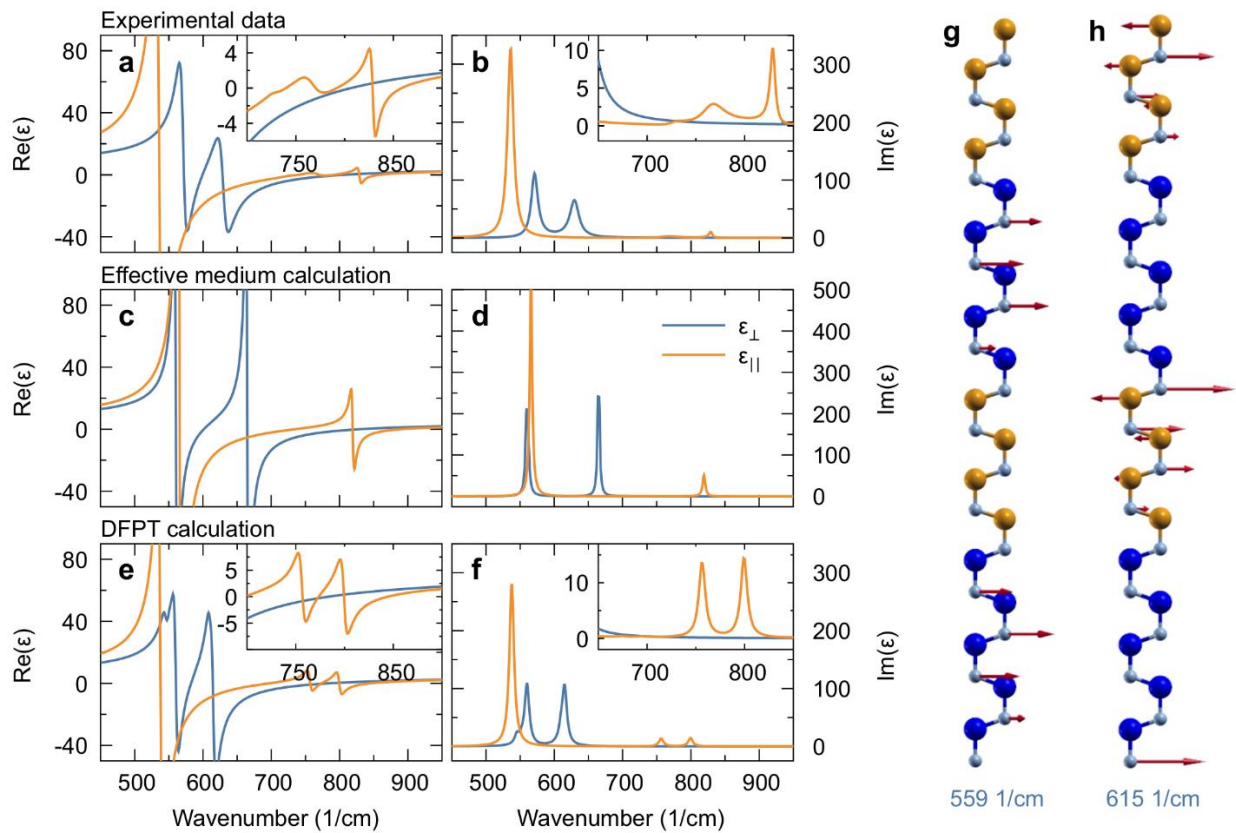

**Figure 3** Dielectric function of the atomic-scale AlN/GaN SL. a-b) measured, c-d) effective medium calculated, and e-f) DFPT calculated real and imaginary parts of the dielectric function of the AlN/GaN SL (Sample A). The in-plane component, $\varepsilon_\perp$, and the out-of-plane component, $\varepsilon_\parallel$, are shown in blue and orange, respectively. g) and h) phonon vibrational patterns associated with the 559-583 cm$^{-1}$ and 615-784 cm$^{-1}$ Reststrahlen bands, respectively, for $\varepsilon_\perp$ shown in e).

**Polaritonic response of the XH**

In order to prove that the XH dielectric function is descriptive of the underlying physics, we show that it accurately predicts the SL's nanophotonic response. If the material acts as a single, continuous medium, the polariton dispersion should be consistent with predictions using the XH dielectric function, in contrast to the effective-medium approximation. To test this, we employ prism coupling in the Otto geometry (see Methods for details)[39,44,45] to measure the dispersion of the polariton modes, as shown in the schematic of Fig. 4a. The experimental dispersion maps of the XH are shown in Fig. 4b, where several dispersing polariton modes are clearly identified. To understand the nature of these modes, thin-film effects and interactions of the XH polaritons with the SiC substrate must be considered (see Supplemental for a detailed discussion).

To corroborate the experimental findings with the XH concept, we calculated the polaritonic dispersion spectra for the Otto geometry using a transfer-matrix method for anisotropic materials[36]. Two complementary models were employed to describe the IR response of the SL structure: (i) the extracted XH dielectric function (Fig. 3a-b), and (ii) an explicit calculation using the bulk dielectric functions for each AlN and GaN thin layer. For very thin layers, the latter model is expected to produce results that are equivalent to the effective-medium theory. The results of the transfer matrix calculations for several cuts through the dispersion (Fig. 4b) are shown together with the respective experimental spectra in Fig. 4c-e. From these comparisons, we find very good agreement with the model using (i) the XH dielectric function (solid orange curve), while (ii) the multilayer calculations (dashed red curve) using the bulk properties of the constituent materials cannot reproduce the observed polaritonic response. These results provide unequivocal evidence of the realization of the XH concept, and therefore, the ability to use atomic-scale manipulation of the vibrational character of phonon modes in SLs to modify the IR dielectric function.

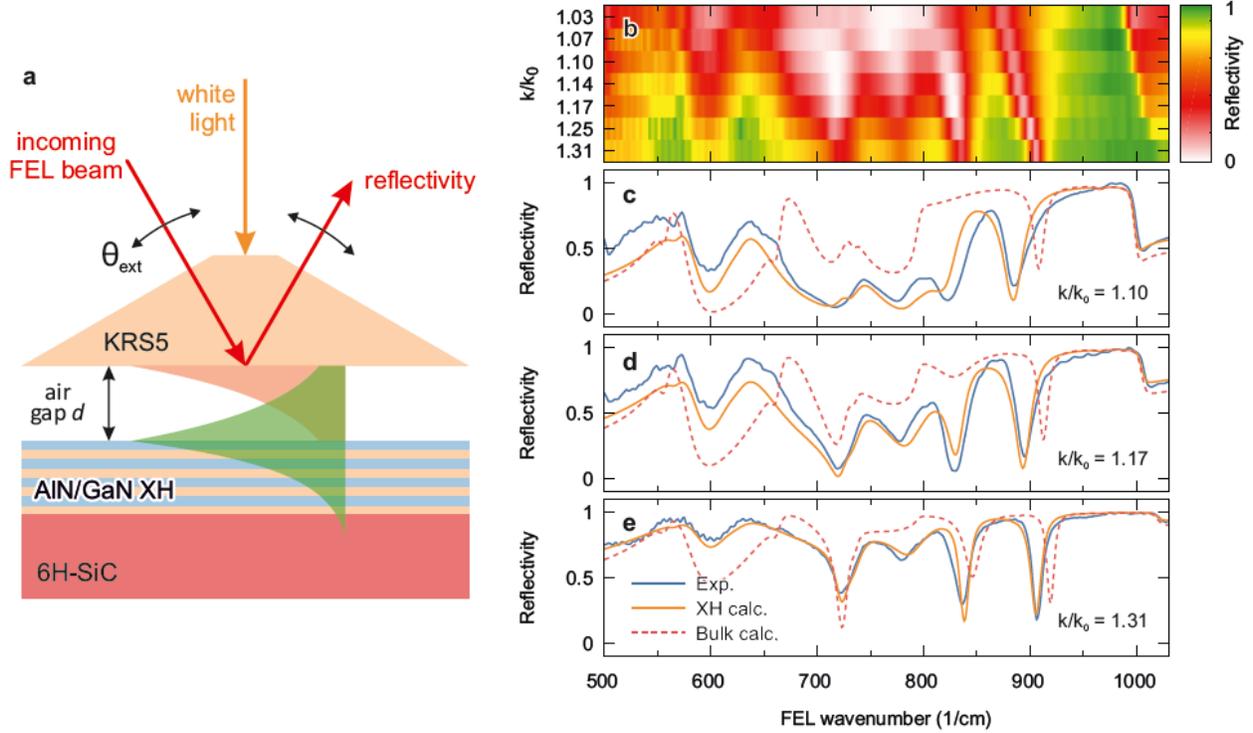

**Figure 4** Polariton dispersion of the XH. (a) Schematic of the Otto-type prism coupling experiment[39] that probes the dispersion of XH polaritons. Total internal reflection at the prism (KRS5, $n = 2.4$) backside launches evanescent waves (red shaded) with large in-plane momenta $k$, tunable via the incidence angle $\theta_{ext}$, that launch evanescent modes in the XH (green shaded) across a well-defined air gap of thickness $d$ determined by white light interferometry, which in turn couple to SPhP modes in the XH. (b) Experimental reflectivity map for a series of normalized wavevectors $k/k_0$, where $k_0$ is the wavevector in vacuum, at fixed air gap thickness $d = 3.0$ μm. SPhP modes appear as minima (red-white shading). (c-e) Experimental (blue) and calculated (solid orange and dashed red) reflectivity spectra for selected momenta.

## Conclusions

We have demonstrated the use of atomic-scale AlN/GaN SLs for creating hybrid polar semiconductors with tunable IR responses. The IR properties of these XH materials are governed by their optic phonons, which, in turn, depend on the SL structural geometry and composition. This implies that the resulting SL does not behave as an effective medium that combines two bulk materials, but rather as a new XH with its own distinct phonon density of states and IR response. Here, we demonstrated tunability of the XH optic phonons based on changes to the constituent layer thicknesses (with frequency shifts >10 cm$^{-1}$). This provides the opportunity to manipulate the IR range over which polariton modes can be supported and to modify the dispersion of the optical constants, without introducing additional sources of optical loss. We showed the AlN/GaN XH has multiple Reststrahlen bands, offering the potential for tailoring the spectral dispersion of both elliptical and hyperbolic spectral regions. A major benefit of the XH approach is that it allows the combination of two materials to achieve a desired IR response, while maintaining their individual mechanical, electrical, and/or opto-electronic functionality. This development could add flexibility in the design of multi-functional nanophotonic components, while keeping optical losses relatively low. For example, one can envision active devices whose electronic properties are dictated by one constituent (e.g. GaN), where the XH polaritonic response is modulated at high speeds through carrier injection[18].

**Methods**

*Sample growth*: We used radio-frequency (RF) plasma-assisted molecular beam epitaxy (MBE) to grow the polar epitaxial heterostructures on 3-inch diameter metal-polar semi-insulating 4H- and 6H-SiC substrates. The substrates were commercially polished using chemical-mechanical polishing to an epi-ready finish and were used as received. The reactive nitrogen was generated using an RF plasma source fed by ultra-high purity N$_2$ which was further purified by an in-line purifier. The Ga and Al fluxes were generated using conventional dual-filament effusion cells. Other details about the MBE growth conditions were given in an earlier publication[46]. X-ray diffraction (XRD) measurements were carried out using a Rigaku system that employs a rotating Cu anode to produce Cu-Ka radiation. AlN/GaN SL thicknesses were estimated by fitting satellite peaks in XRD scans and high resolution cross-sectional scanning transmission electron microscopy (STEM) images. The SL of Sample A was grown on ~50 nm AlN buffer layer. The SL consists of 50 alternating layers of AlN and GaN and was fabricated with a gradient of AlN and GaN thickness across the SiC wafer ranging from about 2-3 nm thick for both the AlN and GaN layers. For Sample B, there are 500 alternating layers of GaN and AlN, with average layer thicknesses of 1.35 +/- 0.13 nm (4.78 monolayers) and 1.17 +/- 0.05 nm (4.13 monolayers), respectively.

*Electron microscopy preparation and imaging*: Cross-sectional STEM specimens were prepared as lift-out sections with an FEI Helios, a focused ion beam scanning electron microscope. Initial cuts were made at 30 kV, and final polishing was performed at 8 kV. STEM examination was carried out using a Nion UltraSTEM200-X operated at an accelerating voltage of 200 kV. Bright field, medium and high-angle annular dark-field images were acquired. The image magnification calibration was verified each session using lattice measurements of evaporated gold nanoparticles.

*Ellipsometry*: Infrared ellipsometry was performed using a J.A. Woolam Mark II IR-VASE spectroscopic ellipsometer. The sample was mounted in the upright position and aligned using the standard four-quadrant detector to be positioned at the center of rotation. The broadband incident light was provided by a SiC glow-bar and was detected using a DLaTGS detector. The incident light was polarized and detected using a polarizer and analyzer combination, and the various ratios of the p- and s-polarization components of the reflected light were plotted as Ψ and Δ, as is standard for ellipsometric measurements. The Ψ and Δ signal were both measured as a function of frequency (250-8000 cm$^{-1}$) at angles of incidence of 45, 55, 65, and 75° using the standard $\theta/2\theta$ geometry. The collected data was modeled using the WVASE software and based on the phonon frequencies extracted from the SHG measurements provided the starting parameters for the best fit. Following least-squares fitting, the resultant dielectric function was extracted.

*Second Harmonic Spectroscopy:* The second-harmonic phonon spectra[37,42] were obtained using an IR free-electron laser (FEL)[41] as a light source which provides widely tunable, ~ 0.5 % bandwidth laser pulses in the mid- to far-IR. Using a noncollinear autocorrelator geometry in reflection, both FEL beams were focused onto the SL wafer in spatial and temporal overlap, incident at angles $\alpha_1^i = 28°$ and $\alpha_2^i = 62°$, respectively. The SHG signal emerges spatially separated from the reflected fundamental beams and was detected using a LN$_2$-cooled mercury cadmium telluride (MCT) detector (InfraRed Associates). Both incoming beams as well as the detected SHG beam were p-polarized. Optical long- and shortpass filters were used in the incoming and detected beams, respectively, to suppress intrinsic higher harmonics of the FEL and scattered fundamental light. Simultaneously, the intensity of the reflected beam incident at $\alpha_2^i = 62°$ was recorded using a pyroelectric detector. Varying the FEL undulator gap allows for tuning of the FEL center frequency $\omega$, thereby providing the SHG and reflectivity spectra shown in Fig. 2. Spectra were recorded at different locations on the SL wafer corresponding to varying absolute and relative thicknesses of the AlN and GaN layers, which were then extracted from XRD measurements for the representative examples shown in Fig. 2. The full data set of the SHG spectra is provided in the Supplemental.

*Otto-geometry measurements:* Prism coupling in the Otto geometry was achieved using a triangular KRS5 ($n \approx 2.4$) prism in total internal reflection which provides the in-plane momenta necessary to couple to SPhPs[39]. The prism was mounted on a motorized holder which allows for a controlled adjustment of the air gap width $d$, which is read out by white light interferometry, see Fig. 4a. The prism-sample assembly itself was mounted on a motorized rotation stage in order to vary the incidence angle $\theta_{ext}$, and consequently the in-plane momentum $k = \omega/c \; n \sin\theta_{ext}$ of the incoming wave, thereby providing the means to excite SPhPs at different points along their dispersion. Scanning the FEL center frequency $\omega$ across the XH's Reststrahlen region while detecting the reflected intensity resulted in the reflectivity spectra shown in Fig. 4. The data were normalized to a reference spectrum taken at large air gaps (d > 100 μm), where no light can be coupled to the SPhP modes. This measurement was repeated at various incidence angles $\theta_{ext}$, and thus in-plane momenta $k$, to map out the polariton dispersion of the XH. Since multiple modes with different critical coupling behavior appear in the XH structure, the data shown in Fig. 4 were recorded at a fixed air gap width of $d = 3.0 \; \mu m$.

*Transfer matrix calculations:* The theoretical reflectivity spectra in Fig. 1 as well as the theoretical Otto geometry spectra in Fig. 4 were acquired using a transfer matrix algorithm specifically accounting for absorptive anisotropic media[36]. The algorithm can treat an arbitrary number of layers, allowing to directly compare spectra for the XH and the explicit multilayer stack using the bulk dielectric functions of SiC, AlN and GaN. The bulk parameters for SiC, AlN and GaN are reported in the Supplemental.

*DFPT*: The Γ-point phonon frequencies were obtained using the first-principles approach as implemented in the Quantum-ESPRESSO package[47]. In this work, we used the generalized gradient approximation (GGA)[48] of Perdew-Burke-Ernzerhof (PBE)[49] for the exchange-correlation functional. The electronic wavefunctions were expanded in the plane wave basis set with a kinetic energy cut-off of 40 Ry. The Brillouin-zone was sampled with a grid of $18 \times 18 \times 2$ k-points according to the Monkhort-Pack method[50]. In order to compare theoretical results with those reported in the experiment, the GaN/AlN heterostructure was created using 5 monolayers of GaN and 4 monolayers of AlN. The geometry of the resultant structure was optimized and the phonon frequencies for the ground-state structure were calculated within Density-Functional Perturbation Theory (DFPT)[51-54]. Originating from long-ranged dipole-dipole interactions, the LO-TO splitting at the zone center ($\vec{q} \to 0$) was taken into account by including the nonanalytical contribution to the dynamical matrix.

## Acknowledgements

D.C.R., C.T.E., J.G.T., I.V., T.R., N.N., A.J.G., D.S.K., N.D.B., M.T.H., R.M.S. and J.D.C. were supported by the Office of Naval Research through the U.S. Naval Research Laboratory and administered by the NRL Nanoscience Institute. J.D.C. and D.C.R. would like to express their sincere gratitude to Dr. Thomas Tiwald of J.A. Woolam, Inc. for his insight and assistance in performing the dielectric function fitting of the SLs explored in this work. C.J.W., N.C.P., I.R. and A.P. would like to thank Wieland Schoellkopf and Sandy Gewinner for operating the IR-FEL. P.D. acknowledges support from NRL through the ONR Summer Faculty Program. Computer resources were provided by the DoD High Performance Computing Modernization Program. I.C., J.W., M.K. acknowledge support from the NRC/ASEE Postdoctoral Fellowship at NRL. C.T.E. acknowledges support from the U.S. Naval Research Laboratory, Karles Fellowship. Microscopy research was performed as part of a user proposal at Oak Ridge National Laboratory's Center for Nanophase Materials Sciences (CNMS), which is a U.S. Department of Energy, Office of Science User Facility at the Oak Ridge National Laboratory.

## Author Contributions

J.D.C., T.L.R., I.V. and J.G.T. originated the concept, while A.P. devised the SHG methods and Otto-configuration experiments. The manuscript was written by D.C.R., C.J.W., J.D.C and A.P., with all authors assisting in the proof-reading and preparation for final submission. D.S.K. grew

the samples, N.N. preformed XRD and AFM characterization, and M.T.H. performed XRD reciprocal space mapping and modeling. J.D.C., I.C. and A.J.G. performed the IR ellipsometry, while D.C.R. performed the least-squares fitting. SHG and Otto configuration measurements were performed by C.J.W., N.C.P., I.R. and A.P. Computations of the Otto-configuration-based reflectance measurements were completed by A.P. and N.P. IR reflection data was collected by C.T.E. and J.G.T. STEM sample preparation and measurement was performed by M.K., J.W., N.D.B., R.M.S., J.A.H. and J.C.I. First principles calculations were undertaken by P.D. and T.L.R. The experimental design and project management were provided by J.D.C. and A.P.**References**

1  Dai, S. *et al.* Subdiffractional focusing and guiding of polaritonic rays in a natural hyperbolic material. *Nat. Commun.* **6**, 6963, doi:10.1038/ncomms7963 (2015).
2  Li, P. N. *et al.* Hyperbolic phonon-polaritons in boron nitride for near-field optical imaging and focusing. *Nat. Commun.* **6**, 7507, doi:10.1038/ncomms8507 (2015).
3  Liu, Z. W., Lee, H., Xiong, Y., Sun, C. & Zhang, X. Far-field optical hyperlens magnifying sub-diffraction-limited objects. *Science* **315**, 1686-1686, doi:10.1126/science.1137368 (2007).
4  Taubner, T., Korobkin, D., Urzhumov, Y., Shvets, G. & Hillenbrand, R. Near-field microscopy through a SiC superlens. *Science* **313**, 1595-1595, doi:10.1126/science.1131025 (2006).
5  Autore, M. *et al.* Boron nitride nanoresonators for phonon-enhanced molecular vibrational spectroscopy at the strong coupling limit. *Light: Science &Amp; Applications* **7**, 17172, doi:10.1038/lsa.2017.172 https://www.nature.com/articles/lsa2017172#supplementary-information (2018).
6  Jackson, S. D. Towards high-power mid-infrared emission from a fibre laser. *Nat Photon* **6**, 423-431 (2012).
7  Adato, R. *et al.* Ultra-sensitive vibrational spectroscopy of protein monolayers with plasmonic nanoantenna arrays. *Proc. Natl. Acad. Sci. U.S.A.* **106**, 19227-19232, doi:10.1073/pnas.0907459106 (2009).
8  Caldwell, J. D. *et al.* Low-Loss, Extreme Subdiffraction Photon Confinement via Silicon Carbide Localized Surface Phonon Polariton Resonators. *Nano Lett.* **13**, 3690-3697, doi:10.1021/nl401590g (2013).
9  Caldwell, J. D. *et al.* Sub-diffractional volume-confined polaritons in the natural hyperbolic material hexagonal boron nitride. *Nat. Commun.* **5**, 5221, doi:10.1038/ncomms6221 (2014).
10  Caldwell, J. D. *et al.* Low-loss, infrared and terahertz nanophotonics using surface phonon polaritons. *Nanophotonics* **4**, 44-68, doi:10.1515/nanoph-2014-0003 (2015).
11  Dai, S. *et al.* Tunable Phonon Polaritons in Atomically Thin van der Waals Crystals of Boron Nitride. *Science* **343**, 1125-1129, doi:10.1126/science.1246833 (2014).
12  Greffet, J. J. *et al.* Coherent emission of light by thermal sources. *Nature* **416**, 61-64, doi:10.1038/416061a (2002).
13  Hillenbrand, R., Taubner, T. & Keilmann, F. Phonon-enhanced light-matter interaction at the nanometre scale. *Nature* **418**, 159-162, doi:10.1038/nature00899 (2002).21